\documentclass[11pt,a4paper,english,superscriptaddress,aps,preprint]{revtex4}
\usepackage{xcolor}
\usepackage{amsfonts}
\usepackage{graphics}
\usepackage{graphicx}
\usepackage{hyperref}
\usepackage{slashed}
\usepackage{amsmath}
\usepackage{amssymb}
\makeatletter
\usepackage{babel}
\newcommand{\bea}{\begin{eqnarray}}
\newcommand{\eea}{\end{eqnarray}}

\begin{document}

\title{On supersymmetric gauge theories with higher derivatives and nonlocal terms in the matter sector}

\author{F. S. Gama}
\email{fisicofabricio@yahoo.com.br}
\affiliation{Departamento de F\'{\i}sica, Universidade Federal da Para\'{\i}ba\\
 Caixa Postal 5008, 58051-970, Jo\~ao Pessoa, Para\'{\i}ba, Brazil}

\author{J. R. Nascimento}
\email{jroberto@fisica.ufpb.br}
\affiliation{Departamento de F\'{\i}sica, Universidade Federal da Para\'{\i}ba\\
 Caixa Postal 5008, 58051-970, Jo\~ao Pessoa, Para\'{\i}ba, Brazil}

\author{A. Yu. Petrov}
\email{petrov@fisica.ufpb.br}
\affiliation{Departamento de F\'{\i}sica, Universidade Federal da Para\'{\i}ba\\
 Caixa Postal 5008, 58051-970, Jo\~ao Pessoa, Para\'{\i}ba, Brazil}

\begin{abstract}
In this work, we consider local and nonlocal higher-derivative generalizations of the super-Chern-Simons theory and four-dimensional SQED. In contrast to previous studies, the models studied here also have higher-derivative terms in the matter sector. 
For these models, we calculate the one-loop superfield effective potential.
\end{abstract}

\maketitle

\section{Introduction}

Historically, higher-derivative theories have been introduced in an attempt to prevent singularities in a classical field theory \cite{BPS}, and to avoid ultraviolet divergences of a quantum field theory \cite{LWS}. In supersymmetric models, higher-derivative theories have been studied within different contexts. For example, the phenomenological implications of a extension of the Minimal Supersymmetric Standard Model with dimensions-five and six operators have been investigated in \cite{ADGT}. In \cite{galileons} supersymmetric versions of cubic and quartic Galileon theories were proposed. Nonlocal higher-derivative extensions for the scalar, super Yang-Mills, and supergravity theories have been constructed in \cite{nonlocal}. In \cite{ghostbusters}, a new mechanism to construct ghost-free higher-derivative models was formulated. Recently, the higher-covariant-derivative regularization has successfully been applied in supersymmetric gauge theories \cite{Stepanyantz}.

The effective potential is an important theoretical tool for studying the ground state of a theory and the phenomena related to it, such as the spontaneous breaking and restoration of symmetries \cite{breaking,restoration}. In the context of higher-derivative superfield theories, the effective potential has been investigated for different models \cite{SHD,GNP,DGNP}. In particular, in Ref. \cite{GGNPS}, the one-loop effective potential has been explicitly calculated for the simplest higher-derivative extension of a abelian gauge superfield theory. In \cite{GNP2} and \cite{GGNPS2}, the effective potential has been studied in higher-derivative gauge superfield theories defined on the $\mathcal{N}=1$ and $\mathcal{N}=2$ three-dimensional superspaces. More recently, a nonlocal higher-derivative extension of the supersymmetric gauge theory was proposed and the one-loop K\"{a}hlerian effective potential has been explicitly calculated for this theory \cite{GNPP}.

One important limitation of the higher-derivative gauge superfield theories studied in \cite{GGNPS,GNP2,GGNPS2,GNPP} is that they do not include higher derivatives in the matter sector. In particular, the four-dimensional theories studied in Refs. \cite{GGNPS,GNPP} also do not contain chiral self-interaction terms. Since these terms give non-trivial contributions to the one-loop superfield effective potential, there is no reason (other than convenience for calculating the one-loop effective potential) to ignore higher-derivative and chiral self-interaction terms in the matter sector of a higher-derivative supersymmetric gauge theory. Thus, the aim of this paper is to formulate higher-derivative or nonlocal gauge covariant terms in the matter sector and to calculate the superfield effective potential at the one-loop level for local and nonlocal higher-derivative generalizations of the super-Chern-Simons theory and SQED by taking into account these new terms in the matter sector. In this regard, our work is a further development of the studies presented in \cite{GGNPS,GNP2,GGNPS2,GNPP}.

This paper is organized as follows. In section II, we formulate a generic higher-derivative Super-Chern-Simons Theory coupled to matter and calculate the one-loop contribution to the superfield effective potential. In section III, we formulate a generic higher-derivative four-dimensional SQED and explicitly calculate the one-loop K\"{a}hlerian effective potential in it. In section IV, we give a short summary of the results obtained and suggest a possible continuation of this study.

\section{Higher-Derivative Super-Chern-Simons Theory}

Our starting point is the following $\mathcal{N}=1$, $d=3$ higher-derivative action for the complex scalar multiplet:
\bea
\label{3dHM}
S_{HM}=\frac{1}{2}\int d^5z\left[\bar{\Phi}\left(f(\Box)D^2+mg(\Box)\right)\Phi+h.c.\right]+\int d^5zV(\bar{\Phi}\Phi) \ ,
\eea
which is invariant under the global transformations $\delta \Phi=iK\Phi$ and $\delta\bar{\Phi}=-iK\bar{\Phi}$. This model was originally proposed in Ref. \cite{GNP} and studied only in the context of local theories. The dimensionless operators $f(\Box)$ and $g(\Box)$ are assumed to be analytical functions of the d'Alembertian operator. Additionally, in order to reproduce the standard action for the complex scalar multiplet, we also suppose that $f(\Box)$ and $g(\Box)$ coincide with the unit operator in some suitable limit.
 
We are interested in coupling of the theory (\ref{3dHM}) to the abelian gauge superfield $A_\alpha$. In order to do this, we will use the identity $\Box=(D^2)^2$ and apply the minimal coupling prescription \cite{GGRS} changing the simple covariant derivative by the gauge covariant one through the rule
\bea
D_\alpha\Phi&\longrightarrow&\nabla_\alpha\Phi\equiv D_\alpha\Phi-iA_\alpha\Phi \ .
\eea
Thus, Eq. (\ref{3dHM}) can be rewritten as
\bea
\label{3dHD}
S_{HM}=\frac{1}{2}\int d^5z\left[\bar{\Phi}\left(f(\nabla^4)\nabla^2+mg(\nabla^4)\right)\Phi+h.c.\right]+\int d^5zV(\bar{\Phi}\Phi) \ .
\eea
Evidently, this gauged model is invariant under the local transformations
\bea
\label{gaugetrans}
[\Phi]^\prime=e^{iK}\Phi \ \ ; \ \ [A_\alpha]^\prime=A_\alpha+D_\alpha K \ .
\eea
Since $A_\alpha$ is a non-dynamical superfield in (\ref{3dHD}), to introduce a consistent dynamics for it we will add to (\ref{3dHD}) the following higher-derivative generalization of the supersymmetric Chern-Simons theory
\bea
\label{HCS}
S_{HCS}=\frac{1}{2e^2}\int d^5zA^\alpha h(\Box)D^\beta D_\alpha A_\beta \ ,
\eea
which is also invariant under the transformations (\ref{gaugetrans}).

Finally, the higher-derivative version of the super-Chern-Simons theory coupled to matter superfields that we will study in this work has the following action
\bea
\label{3dtotalac}
S=S_{HCS}+S_{HM}+S_{GF} \ ,
\eea
where, to perform quantum calculations, we conveniently added the gauge-fixing functional
\bea
\label{gf}
S_{GF}=\frac{1}{2e^2\alpha}\int d^5zA^\alpha h(\Box)D_\alpha D^\beta A_\beta \ .
\eea
In order to carry out the calculation of the one-loop superfield effective potential in three dimensions \cite{FGLNPSS}, we will employ the background field method \cite{DeWitt}. Making the background-quantum splitting $\Phi\rightarrow\Phi+\phi$ in (\ref{3dtotalac}), assuming that the background superfield satisfies the condition $D_\alpha\Phi=0$, and expanding the action to up to the second order in the quantum superfields,  after some tedious but straightforward manipulations we obtain
\bea
\label{s2}
S_2=\int d^5z\left(\frac{1}{2}A^\alpha\hat{H}_\alpha^{\ \beta}A_\beta+A^\alpha\mathcal{F}_\alpha\right)+\frac{1}{2}\int d^5z
\boldsymbol{\phi}^T
\hat{O}
\boldsymbol{\phi} \ ,
\eea
where
\bea
\label{HF}
\hat{H}_\alpha^{\ \beta}&\equiv&\frac{h(\Box)}{e^2}\left(D^\beta D_\alpha+\frac{1}{\alpha}D_\alpha D^\beta\right)+\frac{\left|\Phi\right|^2}{2}\left(-D^\beta D_\alpha+f(\Box)D_\alpha D^\beta\right)\frac{D^2}{\Box}\nonumber\\
&-&\frac{m\left|\Phi\right|^2}{2}\frac{\left[g(\Box)-1\right]}{\Box}D_\alpha D^\beta \ ;\\
\mathcal{F}_\alpha &\equiv&\frac{i}{2}\left\{\Phi f(\Box)D_\alpha\bar{\phi}-\bar{\Phi}f(\Box)D_\alpha\phi+m\Phi\frac{\left[g(\Box)-1\right]}{\Box}D_\alpha D^2\bar{\phi}-m\bar{\Phi}\frac{\left[g(\Box)-1\right]}{\Box}D_\alpha D^2\phi\right\}\nonumber \ ,
\eea
and
\bea
\boldsymbol{\phi}\equiv\left(\begin{array}{c}
\phi \\
\bar\phi
\end{array}\right) \ ; \ \hat{O}\equiv\left(\begin{array}{cc}
\displaystyle V_{\Phi\Phi} & \displaystyle f(\Box)D^2+mg(\Box)+V_{\Phi\bar{\Phi}}\\
\displaystyle f(\Box)D^2+mg(\Box)+V_{\Phi\bar{\Phi}} & \displaystyle V_{\bar{\Phi}\bar{\Phi}}
\end{array}\right). 
\eea
For present purposes, it is useful to diagonalize (\ref{s2}). To do this, let us consider the following nonlocal change of variables \cite{transformation}
\bea
\label{3dchange}
A_\alpha(z)\longrightarrow A_\alpha(z)-\int d^5z^\prime {G_\alpha}^\beta(z,z^\prime)\mathcal{F}_\beta(z^\prime) \ ,
\eea
where ${G_\alpha}^\beta(z,z^\prime)$ is the Green's function of the operator $\hat{H}_\alpha^{\ \beta}$ defined from the equation
\bea
\label{3dgreen}
{G_\alpha}^\beta(z,z^\prime)=\left(AD^\beta D_\alpha+BD_\alpha D^\beta\right)\delta^5(z-z^\prime) \ ,
\eea
where the coefficients $A$ and $B$ are written in the Appendix.

Under the nonlocal transformation (\ref{3dchange}), the functional (\ref{s2}) assumes the diagonalized form
\bea
\label{3ddiagonalized}
S_2=\frac{1}{2}\int d^5zA^\alpha\hat{H}_\alpha^{\ \beta}A_\beta+\frac{1}{2}\int d^5z
\boldsymbol{\phi}^T
\hat{O}
\boldsymbol{\phi}-\int d^5zd^5z^\prime \mathcal{F}^\alpha(z){G_\alpha}^\beta(z,z^\prime)\mathcal{F}_\beta(z^\prime) \ .
\eea
Since (\ref{3dchange}) is merely a shift by a constant, it leaves the integration measure in the path integral invariant. By  integrating  out  the  quantum superfields $A_\alpha$ and $\boldsymbol{\phi}$, we get two contributions to the Euclidean one-loop effective action
\bea
\label{3dgamma}
\Gamma^{(1)}=\Gamma^{(1)}_A+\Gamma^{(1)}_{\boldsymbol{\phi}} \ .
\eea
The first contribution $\Gamma^{(1)}_A$ is given by the trace:
\bea
\label{3dfirsttrace}
\Gamma^{(1)}_A=\frac{1}{2}\textrm{Tr}\ln\hat{H}_\alpha^{\ \beta}&=&\frac{1}{2}\textrm{Tr}\ln\bigg\{\frac{h(\Box)}{e^2}\left(D^\gamma D_\alpha+\frac{1}{\alpha}D_\alpha D^\gamma\right)\bigg\}+\frac{1}{2}\textrm{Tr}\ln\bigg\{{\delta_\gamma}^\beta-\frac{M}{2\Box h(\Box)}D^\beta D_\gamma\nonumber\\
&-&\frac{\alpha Mf(\Box)}{2\Box h(\Box)}D_\gamma D^\beta+\frac{\alpha mM\left[g(\Box)-1\right]}{2\Box^2 h(\Box)}D_\gamma D^\beta D^2\bigg\} \ ,
\eea
where we factored out the inverse of the propagator of $A_\alpha$ and defined $M\equiv\frac{1}{2}e^2\left|\Phi\right|^2$.

Since the first trace does not depend on the background superfield, we can drop it out. The second trace can be simplified if we
assume the Landau gauge $\alpha=0$. Therefore, it follows from (\ref{3dfirsttrace}) that
\bea
\label{3dfirst}
\left.\Gamma^{(1)}_A\right|_{\alpha=0}=\frac{1}{2}\textrm{Tr}\ln\bigg\{{\delta_\gamma}^\beta-\frac{M}{2\Box h(\Box)}D^\beta D_\gamma\bigg\}=\frac{1}{2}\int d^5z\int \frac{d^3k}{(2\pi)^3}\frac{1}{|k|}\arctan\left[\frac{M}{|k|h(-k^2)}\right] \ .
\eea
We now determine the second contribution $\Gamma^{(1)}_{\boldsymbol{\phi}}$. If we impose the Landau gauge, then ${G_\alpha}^\beta(z,z^\prime)=AD^\beta D_\alpha\delta^5(z-z^\prime)$ and the last term in Eq. (\ref{3ddiagonalized}) vanishes due to the identity $D^\alpha D_\beta D_\alpha=0$. Therefore, we can write
\bea
\label{secondtrace0}
\left.\Gamma^{(1)}_{\boldsymbol{\phi}}\right|_{\alpha=0}=-\frac{1}{2}\textrm{Tr}\ln\hat{O}=
-\frac{1}{2}\textrm{Tr}\ln\left(\begin{array}{cc}
\displaystyle0 & \displaystyle f(\Box)D^2\\
\displaystyle f(\Box)D^2 & \displaystyle0
\end{array}\right)-\frac{1}{2}\textrm{Tr}\ln\left[\hat{I}_2+\mathcal{M}\frac{D^2}{\Box f(\Box)}\right]  \ , 
\eea
where we factored out the inverse of the $\boldsymbol{\phi}$-propagator and defined
\bea
\mathcal{M}\equiv\left(\begin{array}{cc}
\displaystyle mg(\Box)+V_{\bar{\Phi}\Phi} & \displaystyle V_{\bar{\Phi}\bar{\Phi}}\\
\displaystyle V_{\Phi\Phi} & \displaystyle mg(\Box)+V_{\bar{\Phi}\Phi}
\end{array}\right) \ .
\eea
Again, we can drop the first trace out and the second one can be evaluated to give 
\bea
\label{3dsecond}
\left.\Gamma^{(1)}_{\boldsymbol{\phi}}\right|_{\alpha=0}=\frac{1}{2}\int d^5z \int \frac{d^3k}{(2\pi)^3}\frac{1}{|k|}\textrm{Tr}\arctan\left[\frac{\mathcal{M}}{|k|f(-k^2)}\right] \ .
\eea
Finally, substituting (\ref{3dfirst}) and (\ref{3dsecond}) into (\ref{3dgamma}), we can infer that the superfield effective potential is given by the expression
\bea
\label{3dint}
K^{(1)}(\Phi,\bar{\Phi})&=&\frac{1}{2}\int \frac{d^3k}{(2\pi)^3}\frac{1}{|k|}\left\{\arctan\left[\frac{M}{|k|h(-k^2)}\right]+\sum_{i=+,-}\arctan\left[\frac{\lambda_i(-k^2)}{|k|f(-k^2)}\right]\right\} \ ,
\eea
where the $\lambda$'s  are the eigenvalues of the matrix $\mathcal{M}$:
\bea
\lambda_\pm(-k^2)=mg(-k^2)+V_{\bar{\Phi}\Phi}\pm\left(V_{\bar{\Phi}\bar{\Phi}}V_{\Phi\Phi}\right)^{\frac{1}{2}} \ .
\eea
The last step of our calculation is to evaluate the integrals in Eq. (\ref{3dint}). However, to evaluate these integrals one must specify $f(-k^2)$, $g(-k^2)$, and $h(-k^2)$. In this work, we will examine two higher-derivative models which lead to an improved ultraviolet behavior of the theory: one local and one nonlocal. 

A simple local higher-derivative model is defined by
\bea
\label{3dmodel1}
f(\nabla^4)=g(\nabla^4)=1-\frac{\nabla^4}{\Lambda_L^2} \ ; \ h(\Box)=1-\frac{\Box}{\Lambda_L^2} \ ,
\eea
where $\Lambda_L$ is the mass scale at which the higher-derivative
contributions begin to be pertinent. It follows from (\ref{3dmodel1}) that
\bea
\label{3dfunctions1}
f(-k^2)=g(-k^2)=h(-k^2)=1+\frac{k^2}{\Lambda_L^2} \ .
\eea
All integrals in this article will be evaluated approximately by employing the strategy of expansion by regions \cite{BS}. Therefore, substituting (\ref{3dfunctions1}) into (\ref{3dint}) and assuming that $\Lambda_L$ is large, we find
\bea
\label{3dfinal1}
K^{(1)}_L(\Phi,\bar{\Phi})&\approx&-\frac{M^2}{16\pi}\left(1+2\frac{M^2}{\Lambda_L^2}+7\frac{M^4}{\Lambda_L^4}+30\frac{M^6}{\Lambda_L^6}+\cdots\right)-
\sum_{i=+,-}\frac{(m+\tilde{\lambda}_i)^2}{16\pi}\bigg[1+2\frac{\tilde{\lambda}_i(m+\tilde{\lambda}_i)}{\Lambda_L^2}\nonumber\\
&+&\frac{\tilde{\lambda}_i(m+\tilde{\lambda}_i)^2(2m+7\tilde{\lambda}_i)}{\Lambda_L^4}+2\frac{\tilde{\lambda}_i(m+\tilde{\lambda}_i)^3(m^2+9m\tilde{\lambda}_i+15\tilde{\lambda}_i^2)}{\Lambda_L^6}+\cdots\bigg] \ ,
\eea 
where $\tilde{\lambda}_\pm=V_{\bar{\Phi}\Phi}\pm\left(V_{\bar{\Phi}\bar{\Phi}}V_{\Phi\Phi}\right)^{\frac{1}{2}}$.

On the other hand, a simple nonlocal model is defined by
\bea
\label{3dmodel2}
f(\nabla^4)=g(\nabla^4)=\exp\left(-\frac{\nabla^4}{\Lambda_{NL}^2}\right) \ ; \ h(\Box)=\exp\left(-\frac{\Box}{\Lambda_{NL}^2}\right) \ ,
\eea
where, similar to the local model, $\Lambda_{NL}$ describes the characteristic energy at which the nonlocal contributions become important. According to (\ref{3dmodel2}), we have
 \bea
\label{3dfunctions2}
f(-k^2)=g(-k^2)=h(-k^2)=\exp\left(\frac{k^2}{\Lambda_{NL}^2}\right) \ .
\eea
Therefore, substituting (\ref{3dfunctions2}) into (\ref{3dint}) and assuming that $\Lambda_{NL}$ is large, we get
\bea
\label{3dfinal2}
K^{(1)}_{NL}(\Phi,\bar{\Phi})&\approx&-\frac{M^2}{16\pi}\left(1+2\frac{M^2}{\Lambda_{NL}^2}+6\frac{M^4}{\Lambda_{NL}^4}+\frac{64}{3}\frac{M^6}{\Lambda_{NL}^6}+\cdots\right)-\\&-&
\sum_{i=+,-}\frac{(m+\tilde{\lambda}_i)^2}{16\pi}\bigg[1+2(m+\tilde{\lambda}_i)\nonumber\\
&\times&\frac{\tilde{\lambda}_i}{\Lambda_{NL}^2}+\frac{\tilde{\lambda}_i(m+\tilde{\lambda}_i)^2(m+6\tilde{\lambda}_i)}{\Lambda_{NL}^4}+\frac{1}{3}\frac{\tilde{\lambda}_i(m+\tilde{\lambda}_i)^3(m^2+23m\tilde{\lambda}_i+64\tilde{\lambda}_i^2)}{\Lambda_{NL}^6}+\cdots\bigg] \ .\nonumber
\eea 
Since $\Lambda_L$ and $\Lambda_{NL}$ are finite physical parameters, the one-loop effective potentials (\ref{3dfinal1}) and (\ref{3dfinal2}) are UV finite. Notice that this finiteness remains even if we set the parameters to be infinitely large $\Lambda_L\rightarrow\infty$ and $\Lambda_{NL}\rightarrow\infty$ while many higher-derivative or nonlocal theories turn out to be divergent in this limit which is equivalent to removing the higher-derivative term. Indeed, such one-loop finiteness is a characteristic feature of the three-dimensional theories. Moreover, we note that the expressions (\ref{3dfinal1}) and (\ref{3dfinal2}) coincide up to the orders  $\Lambda_L^{-2}$ and $\Lambda_{NL}^{-2}$ in the approximations. This coincidence occurs because the operators (\ref{3dmodel1}) and (\ref{3dmodel2}) are identical in this particular order.

\section{Higher-Derivative SQED}

In the present section, we are interested in a more realistic theory. Thus, let us consider the following four-dimensional matter action:
\bea
S_M=\int d^8z\left(\bar{\Phi}_+\Phi_++\bar{\Phi}_-\Phi_-\right)+\left[\int d^6z\left(m\Phi_-\Phi_+ +W\left(\Phi_-\Phi_+ \right)\right)+h.c.\right]\ ,
\eea
which is invariant under the rigid $U(1)$ transformations:
\bea
\label{trans1}
\left[\Phi_+\right]^\prime=e^{i\lambda}\Phi_+ \ \ \ ; \ \ \ \left[\Phi_-\right]^\prime=e^{-i\lambda}\Phi_- \ \ \ \left[\bar{\Phi}_+\right]^\prime=e^{-i\lambda}\bar{\Phi}_+ \ \ \ ; \ \ \ \left[\bar{\Phi}_-\right]^\prime=e^{i\lambda}\bar{\Phi}_- \ .
\eea
A natural higher-derivative generalization of this model is the following
\bea
\label{HM1}
S_{HM}&=&\frac{1}{2}\int d^8z\bigg(\bar{\Phi}_+f_+\left(\Box\right)\Phi_++\overline{(f_+\left(\Box\right)\Phi_+)}\Phi_++\bar{\Phi}_-f_-\left(\Box\right)\Phi_-\nonumber\\
&+&\overline{(f_-\left(\Box\right)\Phi_-)}\Phi_-\bigg)+\left[\int d^6z\left(m\Phi_-g\left(\Box\right)\Phi_+ +W\left(\Phi_-\Phi_+ \right)\right)+h.c.\right]\ ,
\eea
where again $f_+(\Box)$, $f_-(\Box)$, and $g(\Box)$ are dimensionless analytical functions and coincide with the identity in some suitable limit. This model is essentially a two-superfield version of the one proposed in Ref. \cite{DGNP} which has been studied only in the context of nonlocal theories. 

Due to the chirality of the superfields $\Phi_\pm$, we have $\bar{D}^2D^2\Phi_\pm=\Box\Phi_\pm$. Thus, the action (\ref{HM1}) can be rewritten in a more convenient form
\bea
\label{HM2}
S_{HM}&=&\frac{1}{2}\int d^8z\bigg(\bar{\Phi}_+f_+\left(\bar{D}^2D^2\right)\Phi_++\overline{(f_+\left(\bar{D}^2D^2\right)\Phi_+)}\Phi_++\bar{\Phi}_-f_-\left(\bar{D}^2D^2\right)\Phi_-\nonumber\\
&+&\overline{(f_-\left(\bar{D}^2D^2\right)\Phi_-)}\Phi_-\bigg)+\left[\int d^6z\left(m\Phi_-g\left(\bar{D}^2D^2\right)\Phi_+ +W\left(\Phi_-\Phi_+ \right)\right)+h.c.\right] \ .
\eea
In order to extend the transformations (\ref{trans1}) to local $U(1)$ transformations, we define
\bea
\label{trans2}
\left[\Phi_+\right]^\prime=e^{i\Lambda}\Phi_+ \ \ \ ; \ \ \ \left[\Phi_-\right]^\prime=e^{-i\Lambda}\Phi_- \ \ \ \left[\bar{\Phi}_+\right]^\prime=e^{-i\bar{\Lambda}}\bar{\Phi}_+ \ \ \ ; \ \ \ \left[\bar{\Phi}_-\right]^\prime=e^{i\bar{\Lambda}}\bar{\Phi}_- \ ,
\eea
where the local parameter $\Lambda$ is chiral.

To extend (\ref{HM2}) up to a form invariant under (\ref{trans2}), we must put into use the minimal coupling prescription \cite{BK}: 
\bea
D_\alpha\Phi_{\pm}&\longrightarrow&\nabla_\alpha\Phi_\pm\equiv D_\alpha\Phi_\pm\mp i\Gamma_\alpha\Phi_\pm \ \ ; \ \ \Gamma_\alpha\equiv iD_\alpha V \  \ ;\nonumber\\
\bar{D}_{\dot{\alpha}}\Phi_{\pm}&\longrightarrow&\bar{\nabla}_{\dot{\alpha}}\Phi_\pm\equiv \bar{D}_{\dot{\alpha}}\Phi_\pm=0 \ , 
\eea
where the gauge superfield $V$ and the connection $\Gamma_\alpha$ transform as
\bea
\label{trans3}
\left[V\right]^\prime=V+i(\bar{\Lambda}-\Lambda) \ \ ; \ \ \left[\Gamma_\alpha\right]^\prime=\Gamma_\alpha+D_\alpha\Lambda \ .
\eea
Therefore, Eq. (\ref{HM2}) can be rewritten as
\bea
\label{GHM}
S_{HM}&=&\frac{1}{2}\int d^8z\bigg(\bar{\Phi}_+e^Vf_+\left(\bar{\nabla}^2\nabla^2\right)\Phi_++\overline{(f_+(\bar{\nabla}^2\nabla^2)\Phi_+)}e^V\Phi_++\bar{\Phi}_-e^{-V}f_-\left(\bar{\nabla}^2\nabla^2\right)\Phi_-\nonumber\\
&+&\overline{(f_-(\bar{\nabla}^2\nabla^2)\Phi_-)}e^{-V}\Phi_-\bigg)+\left[\int d^6z\left(m\Phi_-g\left(\bar{\nabla}^2\nabla^2\right)\Phi_+ +W\left(\Phi_-\Phi_+ \right)\right)+h.c.\right] \ .
\eea
This model is invariant under the combined transformations (\ref{trans2}) and (\ref{trans3}). Notice that we introduced the factor $\exp(V)$ to change a $\Lambda$ representation to a $\bar{\Lambda}$ representation of the group \cite{GGRS}.  

Since $V$ has no kinetic term in (\ref{GHM}), we will add to (\ref{GHM}) the following higher-derivative generalization of the supersymmetric abelian gauge theory
\bea
S_{HG}=\frac{1}{16e^2}\left[\int d^6zW^\alpha h\left(\Box\right)W_\alpha+\int d^6\bar{z}\bar{W}^{\dot{\alpha}} h\left(\Box\right)\bar{W}_{\dot{\alpha}}\right]\ ,
\eea
where the superfield strengths are expressed in terms of the gauge superfield as
\bea
W_\alpha=i\bar{D}^2D_\alpha V \ \ ; \ \ \bar{W}_{\dot{\alpha}}=-iD^2\bar{D}_{\dot{\alpha}} V \ .
\eea
Finally, the higher-derivative version of the SQED that we will study in this paper is given by
\bea
\label{totalaction}
S=S_{HG}+S_{HM} \ .
\eea
Here, our goal is to calculate the one-loop correction to the K\"{a}hlerian effective potential \cite{BKY}. Thus, as we have done in the last section, we will expand (\ref{totalaction}) around background superfields: 
\bea
\Phi_+\rightarrow\Phi_++\phi_+ \ \ ; \ \ \Phi_-\rightarrow\Phi_-+\phi_- \ .
\eea
We will assume that the background superfields are subject to the constraints $D_\alpha\Phi_+=0$ and $D_\alpha\Phi_-$=0, and keep only quadratic terms in the quantum fluctuations $\phi_+$ and $\phi_-$. Therefore, after a lengthy algebra, we find
\bea
\label{S21}
S_2&=&-\frac{1}{8e^2}\int d^8z V\Box h\left(\Box\right)\Pi_{\frac{1}{2}}V+\frac{1}{2}\int d^8z\Big\{\left|\Phi_+\right|^2V\left[\Pi_{\frac{1}{2}}+f_+(\Box)\Pi_0\right]V+2\bar{\Phi}_+Vf_+(\Box)\phi_+\nonumber\\
&+&2\Phi_+Vf_+(\Box)\bar{\phi}_++2\bar{\phi}_+f_+(\Box)\phi_+\Big\}+\frac{1}{2}\int d^8z\Big\{\left|\Phi_-\right|^2V\left[\Pi_{\frac{1}{2}}+f_-(\Box)\Pi_0\right]V\nonumber\\
&-&2\bar{\Phi}_-Vf_-(\Box)\phi_--2\Phi_-Vf_-(\Box)\bar{\phi}_-+2\phi_-f_-(\Box)\bar{\phi}_-\Big\}-m\int d^8z\Big\{\Phi_-\Phi_+V\nonumber\\
&\times&\frac{g(\Box)-1}{\Box}D^2V+\Phi_-V\frac{g(\Box)-1}{\Box}D^2\phi_+-\Phi_+\phi_-\frac{g(\Box)-1}{\Box}D^2V+h.c.\Big\}+\Big\{\int d^6z\Big[m\phi_-\nonumber\\
&\times&g(\Box)\phi_++\frac{1}{2}\frac{\partial^2 W}{\partial\Phi^2_+}\phi_+^2+\frac{1}{2}\frac{\partial^2 W}{\partial\Phi^2_-}\phi_-^2+\frac{\partial^2 W}{\partial\Phi_+\partial\Phi_-}\phi_+\phi_-\Big]+h.c.\Big\} \ ,
\eea
where $\Pi_{\frac{1}{2}}$ and $\Pi_{0}$ are the transverse and longitudinal projection operators, which are defined as
\bea
\Pi_{1/2}=-\frac{D^\alpha\bar D^2D_\alpha}{\Box}\ ; \ \Pi_{0+}=\frac{\bar D^2 D^2}{\Box} \ ; \ \Pi_{0-}=\frac{D^2\bar D^2}{\Box} \ ; \ \Pi_{0}=\Pi_{0+}+\Pi_{0-} \ .
\eea
At this step of our calculation, we find more advantageous to work with unconstrained superfields than chiral superfields. For this reason, we will write the quantum antichiral and chiral superfields as $\phi_\pm=\bar{D}^2\psi_\pm$ and $\bar{\phi}_\pm=D^2\bar{\psi}_\pm$, where $\psi_\pm$ and $\bar{\psi}_\pm$ are free of differential constraints \cite{GRU}. However, this replacement introduces a new gauge
symmetry, namely $\delta\psi_\pm=\bar{D}^{\dot{\alpha}}\bar{\omega}_{\pm\dot{\alpha}}$ and $\delta\bar{\psi}_\pm=D^{\alpha}\omega_{\pm\alpha}$. Therefore, in order to fix this gauge invariance and the one (\ref{trans3}), we will add to (\ref{S21}) the following gauge-fixing functionals
\bea
\label{GF1}
S_{GF1}&=&-\frac{1}{8e^2\alpha}\int d^8z V\Box h(\Box)\Pi_0 V \ ;\\
\label{GF2}
S_{GF2}&=&\int d^8z \bar{\psi}_+f_+(\Box)\left(\bar{D}^2D^2-D^\alpha\bar{D}^2D_\alpha\right)\psi_+ \ ;\\
\label{GF3}
S_{GF3}&=&\int d^8z \bar{\psi}_-f_-(\Box)\left(\bar{D}^2D^2-D^\alpha\bar{D}^2D_\alpha\right)\psi_- \ .
\eea
Therefore, it follows from (\ref{S21}-\ref{GF3}) that
\bea
\label{S22}
\tilde{S}_2\equiv S_2+S_{GF1}+S_{GF2}+S_{GF3}=\int d^8z\left(\frac{1}{2}V\hat{H}V+V\mathcal{F}\right)+\frac{1}{2}\int d^8z
\boldsymbol{\psi}^T
\hat{O}
\boldsymbol{\psi} \ ,
\eea
where
\bea
\hat{H}&\equiv&-\frac{1}{4e^2}\Box h\left(\Box\right)\left[\Pi_{\frac{1}{2}}+\frac{1}{\alpha}\Pi_{0}\right]+\left(\left|\Phi_+\right|^2+\left|\Phi_-\right|^2\right)\Pi_{\frac{1}{2}}+\Big(\left|\Phi_+\right|^2f_+(\Box)\nonumber\\
&+&\left|\Phi_-\right|^2f_-(\Box)\Big)\Pi_{0}-2m\Phi_-\Phi_+\left[g(\Box)-1\right]\frac{D^2}{\Box}-2m\bar{\Phi}_-\bar{\Phi}_+\left[g(\Box)-1\right]\frac{\bar{D}^2}{\Box}\\
\mathcal{F}&\equiv&\bar{\Phi}_+f_+(\Box)\bar{D}^2\psi_+-\bar{\Phi}_-f_-(\Box)\bar{D}^2\psi_--m\Phi_-\left[g(\Box)-1\right]\Pi_{0-}\psi_+\nonumber\\
&+&m\Phi_+\left[g(\Box)-1\right]\Pi_{0-}\psi_-+h.c. \ ,
\eea
and
\bea
\boldsymbol{\psi}\equiv\left(\begin{array}{c}
\psi_+ \\
\psi_- \\
\bar\psi_+ \\
\bar\psi_-
\end{array}\right) \ ; \ \hat{O}\equiv\left(\begin{array}{cc}
\displaystyle\hat{W}\bar{D}^2 & \displaystyle\Box\hat{F}\\
\displaystyle\Box\hat{F} & \displaystyle\hat{\overline{W}}D^2
\end{array}\right) \ . 
\eea
Lastly, the matrices $\hat{W}$ and $\hat{F}$ are defined as
\bea
\hat{F}\equiv\left(\begin{array}{cc}
\displaystyle f_+(\Box) & \displaystyle 0\\
\displaystyle 0 & \displaystyle f_-(\Box)
\end{array}\right) \ ; \ \hat{W}\equiv\left(\begin{array}{cc}
\displaystyle\frac{\partial^2W}{\partial\Phi_+^2} & \displaystyle mg(\Box)+\frac{\partial^2W}{\partial\Phi_+\partial\Phi_-} \\
\displaystyle mg(\Box)+\frac{\partial^2W}{\partial\Phi_+\partial\Phi_-} & \displaystyle\frac{\partial^2W}{\partial\Phi_-^2}
\end{array}\right) \ .
\eea
The mixing terms between the quantum superfields $V$ and $\boldsymbol{\psi}$ can be eliminated by the following nonlocal change of variables in the path integral:
\bea
\label{change}
V(z)\longrightarrow V(z)-\int d^8z^\prime G(z,z^\prime)\mathcal{F}(z^\prime) \ ,
\eea
where $G(z,z^\prime)$ is the Green's function of the operator $\hat{H}=\hat{H}(z)$, namely $\hat{H}G(z,z^\prime)=\delta^8(z-z^\prime)$. This equation has the solution
\bea
\label{green}
G(z,z^\prime)=\left(X\Pi_{\frac{1}{2}}+Y\Pi_0+Z D^2+\bar{Z}\bar{D}^2\right)\delta^8(z-z^\prime) \ .
\eea
The coefficients $X$, $Y$, and $Z$ are written in the Appendix.

Therefore, after the change of variables (\ref{change}), the $\tilde{S}_2$ can be put in the diagonalized form
\bea
\label{S23}
\tilde{S}_2=\frac{1}{2}\int d^8zV\hat{H}V+\frac{1}{2}\int d^8z
\boldsymbol{\psi}^T
\hat{O}
\boldsymbol{\psi}-\int d^8zd^8z^\prime\mathcal{F}(z)G(z,z^\prime)\mathcal{F}(z^\prime) \ .
\eea
From $\tilde{S}_2$, we can compute the Euclidean one-loop effective action by formal integrating out the superfields $V$ and $\boldsymbol{\psi}$. Therefore, we arrive at
\bea
\label{4dgamma}
\Gamma^{(1)}=\Gamma^{(1)}_V+\Gamma^{(1)}_{\boldsymbol{\psi}} \ .
\eea 
The first contribution $\Gamma^{(1)}_V$ is given by the trace:
\bea
\label{firsttrace}
\Gamma^{(1)}_V=-\frac{1}{2}\textrm{Tr}\ln\hat{H}&=&-\frac{1}{2}\textrm{Tr}\ln\bigg\{-\frac{1}{4e^2}\Box h\left(\Box\right)\left[\Pi_{\frac{1}{2}}+\frac{1}{\alpha}\Pi_{0}\right]+\frac{1}{4e^2}M\Pi_{\frac{1}{2}}+\frac{1}{4e^2}\tilde{f}(\Box)\Pi_{0}\nonumber\\
&-&\frac{2m}{4e^2}\tilde{g}(\Box)\frac{D^2}{\Box}-\frac{2m}{4e^2}\tilde{\overline{g}}(\Box)\frac{\bar{D}^2}{\Box}\bigg\} \ ,
\eea
where we introduced the definitions
\bea
M\equiv 4e^2\left(\left|\Phi_+\right|^2+\left|\Phi_-\right|^2\right) \ ; \ \tilde{f}(\Box)\equiv 4e^2\left(\left|\Phi_+\right|^2f_+(\Box)+\left|\Phi_-\right|^2f_-(\Box)\right) \ ;\nonumber\\
\tilde{g}(\Box)\equiv 4e^2\Phi_+\Phi_-\left[g(\Box)-1\right] \ ; \ \tilde{\overline{g}}(\Box)\equiv 4e^2\bar{\Phi}_+\bar{\Phi}_-\left[g(\Box)-1\right] \ .
\eea
We can factor out the inverse of the $V$-propagator, which is independent of the background superfields, and subsequently drop it. Therefore, it follows from (\ref{firsttrace}) that
\bea
\Gamma^{(1)}_V=-\frac{1}{2}\textrm{Tr}\ln\bigg\{1-\frac{M}{\Box h(\Box)}\Pi_{\frac{1}{2}}-\frac{\alpha\tilde{f}(\Box)}{\Box h(\Box)}\Pi_{0}+\frac{2m\alpha\tilde{g}(\Box)}{\Box h(\Box)}\frac{D^2}{\Box}+\frac{2m\alpha\tilde{\overline{g}}(\Box)}{\Box h(\Box)}\frac{\bar{D}^2}{\Box}\bigg\} \ .
\eea
This trace assumes its simplest form in the Landau gauge $\alpha=0$. Therefore, in this particular gauge, we find
\bea
\label{4dfirst}
\left.\Gamma^{(1)}_V\right|_{\alpha=0}=-\frac{1}{2}\textrm{Tr}\ln\left[1-\frac{M}{\Box h(\Box)}\Pi_{\frac{1}{2}}\right]=-\int d^8z\int \frac{d^4p}{(2\pi)^4}\frac{1}{p^2}\ln\left[1+\frac{M}{p^2 h(-p^2)}\right] \ .
\eea
Let us move on to the calculation of the second contribution $\Gamma^{(1)}_{\boldsymbol{\psi}}$. Notice that $G(z,z^\prime)=X\Pi_{\frac{1}{2}}\delta^8(z-z^\prime)$ in the Landau gauge. Since the Green's function is transversal in this gauge, the last term in Eq. (\ref{S23}) vanishes. Therefore, we can write
\bea
\label{secondtrace}
\left.\Gamma^{(1)}_{\boldsymbol{\psi}}\right|_{\alpha=0}=-\frac{1}{2}\textrm{Tr}\ln\hat{O}=-\frac{1}{2}\textrm{Tr}\ln\left(\begin{array}{cc}
\displaystyle\hat{W}\bar{D}^2 & \displaystyle\Box\hat{F}\\
\displaystyle\Box\hat{F} & \displaystyle\hat{\overline{W}}D^2
\end{array}\right) \ . 
\eea
Again, we can factor out the inverse of the $\boldsymbol{\psi}$-propagator and subsequently drop it out (\ref{secondtrace}). Thus, we can rewrite (\ref{secondtrace}) as
\bea
\left.\Gamma^{(1)}_{\boldsymbol{\psi}}\right|_{\alpha=0}=-\frac{1}{2}\textrm{Tr}\ln\left[\hat{I}_4+\left(\begin{array}{cc}
\displaystyle0 & \displaystyle\frac{\hat{F}^{-1}\hat{\overline{W}}}{\Box}D^2\\
\displaystyle\frac{\hat{F}^{-1}\hat{W}}{\Box}\bar{D}^2 & \displaystyle0
\end{array}\right)\right]  \ . 
\eea
Only the even powers in the expansion of the logarithm give non-vanishing contributions to the trace. Therefore, we can show that
\bea
\left.\Gamma^{(1)}_{\boldsymbol{\psi}}\right|_{\alpha=0}&=&-\frac{1}{4}\textrm{Tr}\ln\left[\hat{I}_4-\left(\begin{array}{cc}
\displaystyle\frac{\hat{F}^{-1}\hat{\overline{W}}\hat{F}^{-1}\hat{W}}{\Box}\Pi_{0-} & \displaystyle0\\
\displaystyle0 & \displaystyle\frac{\hat{F}^{-1}\hat{W}\hat{F}^{-1}\hat{\overline{W}}}{\Box}\Pi_{0+}
\end{array}\right)\right]\nonumber\\
&=&-\frac{1}{4}\textrm{Tr}\ln\left(\hat{I}_2-\frac{\hat{F}^{-1}\hat{\overline{W}}\hat{F}^{-1}\hat{W}}{\Box}\Pi_{0-}\right)-\frac{1}{4}\textrm{Tr}\ln\left(\hat{I}_2-\frac{\hat{F}^{-1}\hat{W}\hat{F}^{-1}\hat{\overline{W}}}{\Box}\Pi_{0+}\right) \ .
\eea
These traces can be evaluated to give
\bea
\left.\Gamma^{(1)}_{\boldsymbol{\psi}}\right|_{\alpha=0}=\frac{1}{2}\int d^8z\int \frac{d^4p}{(2\pi)^4}\frac{1}{p^2}\textrm{Tr}\ln\left(\hat{I}_2+\frac{\hat{F}^{-1}\hat{\overline{W}}\hat{F}^{-1}\hat{W}}{p^2}\right) \ .
\eea
This integral is rather complicated. In order to obtain clear analytical results, we will assume that $m=0$ and
\bea
f_+(\bar{\nabla}^2\nabla^2)=f_-(\bar{\nabla}^2\nabla^2)\equiv f(\bar{\nabla}^2\nabla^2) \ .
\eea
Therefore, it follows that
\bea
\label{4dsecond}
\left.\Gamma^{(1)}_{\boldsymbol{\psi}}\right|_{\alpha=0}=\frac{1}{2}\int d^8z\int \frac{d^4p}{(2\pi)^4}\frac{1}{p^2}\textrm{Tr}\ln\left(\hat{I}_2+\frac{\hat{\overline{W}}\hat{W}|_{m=0}}{p^2f^2(-p^2)}\right) \ .
\eea
Finally, substituting (\ref{4dfirst}) and (\ref{4dsecond}) into (\ref{4dgamma}), we can infer that the one-loop correction to the K\"{a}hler effective potential is given by
\bea
\label{kahler}
K^{(1)}(\Phi,\bar{\Phi})&=&\int \frac{d^4p}{(2\pi)^4}\frac{1}{p^2}\left\{-\ln\left[1+\frac{M}{p^2h(-p^2)}\right]+\frac{1}{2}\sum_{i=+,-}\ln\left[1+\frac{\lambda_i}{p^2f^2(-p^2)}\right]\right\} \ ,
\eea
where the $\lambda$'s are the eigenvalues of the matrix $\hat{\overline{W}}\hat{W}|_{m=0}$ and they are given by
\bea
\lambda_\pm&=&\frac{1}{2}\Bigg\{\left|\frac{\partial^2W}{\partial\Phi_+^2}\right|^2+2\left|\frac{\partial^2W}{\partial\Phi_+\partial\Phi_-}\right|^2+\left|\frac{\partial^2W}{\partial\Phi_-^2}\right|^2\pm \Bigg[\left(\left|\frac{\partial^2W}{\partial\Phi_+^2}\right|^2+\left|\frac{\partial^2W}{\partial\Phi_-^2}\right|^2\right)^2\nonumber\\
&+&4\left|\frac{\partial^2W}{\partial\Phi_+\partial\Phi_-}\frac{\partial^2\bar{W}}{\partial\bar{\Phi}_+^2}+\frac{\partial^2\bar{W}}{\partial\bar{\Phi}_+\partial\bar{\Phi}_-}\frac{\partial^2W}{\partial\Phi_-^2}\right|^2\Bigg]^{\frac{1}{2}}\Bigg\} \ .
\eea
In the same manner as the previous section, it is necessary to specify the functions $f(-p^2)$ and $h(-p^2)$ in order to evaluate the integrals in (\ref{kahler}). Thus, again, we will consider one local and one nonlocal higher-derivative model.

The local higher-derivative model is described by
\bea
\label{model1}
f(\bar{\nabla}^2\nabla^2)=1-\frac{\bar{\nabla}^2\nabla^2}{\Lambda_L^2} \ ; \ h(\Box)=1-\frac{\Box}{\Lambda_L^2} \ .
\eea
Thus, we can infer that
\bea
\label{functions1}
f(-p^2)=h(-p^2)=1+\frac{p^2}{\Lambda_L^2} \ .
\eea
Therefore, replacing these functions in (\ref{kahler}) and assuming that $\Lambda_L$ is large, we obtain
\bea
\label{final1}
K^{(1)}_L(\Phi,\bar{\Phi})&\approx&\frac{M}{16\pi^2}\bigg\{-1+\ln\left(\frac{M}{\Lambda_L^2}\right)+\frac{M}{2\Lambda_L^2}\left[1+2\ln\left(\frac{M}{\Lambda_L^2}\right)\right]+\frac{M^2}{6\Lambda_L^4}\left[10+12\ln\left(\frac{M}{\Lambda_L^2}\right)\right]\nonumber\\
&+&\frac{M^3}{12\Lambda_L^6}\left[59+60\ln\left(\frac{M}{\Lambda_L^2}\right)\right]+\cdots\bigg\}-\sum_{i=+,-}\frac{\lambda_i}{32\pi^2}\bigg[\ln\left(\frac{\lambda_i}{\Lambda^2_L}\right)+\frac{\lambda_i}{6\Lambda_L^2}\bigg[13\nonumber\\
&+&12\ln\left(\frac{\lambda_i}{\Lambda_L^2}\right)\bigg]+\frac{\lambda_i^2}{20\Lambda_L^4}\left[193+140\ln\left(\frac{\lambda_i}{\Lambda_L^2}\right)\right]+\frac{\lambda_i^3}{84\Lambda_L^6}\bigg[3825+2520\nonumber\\
&\times&\ln\left(\frac{\lambda_i}{\Lambda_L^2}\right)\bigg]+\cdots\bigg] \ .
\eea
On the other hand, the nonlocal higher-derivative model is described by
\bea
\label{model2}
f(\bar{\nabla}^2\nabla^2)=\exp\left(-\frac{\bar{\nabla}^2\nabla^2}{\Lambda^2_{NL}}\right) \ ; \ h(\Box)=\exp\left(-\frac{\Box}{\Lambda^2_{NL}}\right) \ .
\eea
Evidently, Eq. (\ref{model2}) implies that
 \bea
\label{functions2}
f(-p^2)=h(-p^2)=\exp\left(\frac{p^2}{\Lambda_{NL}^2}\right) \ .
\eea
Therefore, replacing these functions in (\ref{kahler}) and assuming that $\Lambda_{NL}$ is large, we have
\bea
\label{final2}
K^{(1)}_{NL}(\Phi,\bar{\Phi})&\approx&\frac{M}{16\pi^2}\bigg\{\ln\left(\frac{M}{\Lambda_{NL}^2 e^{1-\gamma}}\right)+\frac{M}{\Lambda_{NL}^2}\ln\left(\frac{2M}{\Lambda_{NL}^2 e^{1-\gamma}}\right)+\frac{M^2}{4\Lambda_{NL}^4}\left[-1+6\ln\left(\frac{3M}{\Lambda_{NL}^2 e^{1-\gamma}}\right)\right]\nonumber\\
&+&\frac{4M^3}{9\Lambda_{NL}^6}\left[-2+6\ln\left(\frac{4M}{\Lambda_{NL}^2 e^{1-\gamma}}\right)\right]+\cdots\bigg\}-\sum_{i=+,-}\frac{\lambda_i}{32\pi^2}\bigg[\ln\left(\frac{2\lambda_i}{\Lambda_{NL}^2 e^{1-\gamma}}\right)+\frac{2\lambda_i}{\Lambda_{NL}^2}\nonumber\\
&\times&\ln\left(\frac{4\lambda_i}{\Lambda_{NL}^2 e^{1-\gamma}}\right)+\frac{\lambda_i^2}{\Lambda_{NL}^4}\left[-1+6\ln\left(\frac{6\lambda_i}{\Lambda_{NL}^2 e^{1-\gamma}}\right)\right]+\frac{32\lambda_i^3}{288\Lambda_{NL}^6 }\bigg[-2+6\nonumber\\
&\times&\ln\left(\frac{8\lambda_i}{\Lambda_{NL}^2 e^{1-\gamma}}\right)\bigg]+\cdots\bigg] \ .
\eea 
Like the one-loop effective potentials obtained in the previous section, the K\"{a}hler effective potentials (\ref{final1}) and (\ref{final2}) are also UV-finite. However, when we set the parameters to be infinitely large $\Lambda_L\rightarrow\infty$ and $\Lambda_{NL}\rightarrow\infty$, such finiteness ceases to exist because of the leading term of the potentials. We know this must be so, because in the limits $\Lambda_L\rightarrow\infty$ and $\Lambda_{NL}\rightarrow\infty$ both potentials (\ref{final1}) and (\ref{final2}) must agree with the one for the standard SQED.

\section{Summary}

We considered the higher-derivative/nonlocal extensions of supergauge theories where, unlike previous papers on such theories \cite{GGNPS,GNP2,GGNPS2,GNPP}, the higher derivatives or nonlocality are implemented not only in the gauge sector, but also in the matter sector. It is important to note that within our approach, these kinds of theories are treated within the same methodology. Effectively, we introduced a new class of higher-derivative/nonlocal Abelian supergauge theories and new class of gauge-matter couplings.

We performed the one-loop calculations in these theories with use of the functional supertrace approach and explicitly demonstrated that this approach can be applied to these theories with the same degree of success that to other supergauge theories. In three-dimensional case, the result continues to be finite even when the characteristic scale $\Lambda_L$ (or $\Lambda_{NL}$) goes to infinity, as it must be, since the one-loop effective action is finite in three-dimensional theories. This is not so in four-dimensional case since the one-loop effective potential in the usual SQED diverges \cite{GRU,SQED}, and the higher-derivative/nonlocal terms in various field theory models clearly play the role of the regularization.

The net result of our paper consists in formulating of new gauge-matter couplings. Therefore, it is natural to expect that these couplings can be generalized to other theories, especially to those ones interesting from the phenomenological viewpoint, and  to various effective theories. The advantage of such theories consists in the fact that they are finite and ghost-free. We plan to study phenomenological impacts of new couplings in our next papers.

\vspace*{4mm}

{\bf APPENDIX}

\vspace*{2mm}

Below  the coefficients of the Green's functions (\ref{3dgreen}) and (\ref{green}) are listed:
\bea
A&=&-\frac{e^2}{2}\frac{e^2|\Phi|^2 D^2+2\Box h(\Box)}{e^4|\Phi|^4\Box-4\Box^2 h^2(\Box)} \ ;\\
B&=&\frac{e^2\alpha}{2}\frac{\alpha e^2|\Phi|^2f(\Box)D^2-2\Box h(\Box)+\alpha me^2|\Phi|^2\left[g(\Box)-1\right]}{\alpha^2 e^4|\Phi|^4\Box f^2(\Box)-\left\{2\Box h(\Box)-\alpha me^2|\Phi|^2\left[g(\Box)-1\right]\right\}^2} \ ;\\
X&=&\frac{4e^2}{-\Box h(\Box)+M} \ ;\\
Y&=&\frac{4e^2\alpha\Box\left[-\Box h(\Box)+\alpha\tilde{f}(\Box)\right]}{\Box\left[-\Box h(\Box)+\alpha\tilde{f}(\Box)\right]^2-4\alpha^2m^2\tilde{g}(\Box)\tilde{\overline{g}}(\Box)} \ ;\\
Z&=&\frac{2m\alpha\tilde{g}(\Box)}{\Box\left(-\Box h(\Box)+\alpha\tilde{f}(\Box)\right)}Y \ .
\eea

\vspace{5mm}

{\bf Acknowledgments.} The work by A. Yu. P. has been partially supported by the
CNPq project No. 301562/2019-9.

\end{document}